\documentclass{80SA}
\usepackage{times}

\title{Antiferromagnetism and Superconductivity in CeRhIn$_5$}

\author{Georg \textsc{Knebel}\thanks{E-mail: georg.knebel@cea.fr}, Jonathan \textsc{Buhot}, Dai \textsc{Aoki},   Gerard \textsc{Lapertot}, Stephane \textsc{Raymond}, Eric \textsc{Ressouche}, and Jacques \textsc{Flouquet}} %\\
\inst{INAC/SPSMS, CEA Grenoble, 17 rue des Martyrs, 38054 Grenoble Cedex 9, France%\\
}
\abst{
We discuss recent results on the heavy fermion superconductor CeRhIn$_5$ which presents ideal conditions to study the strong coupling between the suppression of antiferromagnetic order and the appearance of unconventional superconductivity. The appearance of superconductivity as function of pressure is strongly connected to the suppression of the magnetic order. Under magnetic field, the re-entrance of magnetic order inside the superconducting state shows that antiferromagnetism nucleates in the vortex cores. The suppression of antiferromagnetism in CeRhIn$_5$ by Sn doping is compared to that under hydrostatic pressure.}

\kword{unconventional superconductivity, antiferromagnetism, quantum criticality, CeRhIn$_5$}

\begin{document}
\maketitle

\section{Introduction} 
The interplay between rivaling ground states and the appearance of unconventional superconductivity 
 is a central issue in physics of strongly correlated electron systems including heavy fermion metals, but also \cite{Mathur1998}, high-$T_c$ cuprates \cite{Sachdev2010}, organics charge transfer salts \cite{Bourbonnais2008}, and the recently discovered iron pnictides \cite{Paglione2010}. In all these examples unconventional superconductivity appears in the vicinity of a magnetically ordered state and strong deviations from the Landau Fermi liquid description of a metal appears due to the importance of quantum fluctuations. Quantum criticality is the driving mechanism for unconventional superconductivity. 
 
The competition between antiferromagnetic order and unconventional superconductivity has been intensively studied in Ce based heavy fermion systems. A breakthrough was the discovery of the heavy fermion family Ce$M$In$_5$. The crystal structure of these materials is tetragonal and it is build of planes of CeIn$_3$ and planes of $M$In$_2$ ($M$ = Rh, Co, or Ir) which are stacked sequently along the $c$ direction. The cubic CeIn$_3$ itself is an antiferromagnet and superconductivity appears under high pressure around $p = 2.5$ GPa at temperatures below $T_c = 200$~mK \cite{Mathur1998,Knebel2001}. The introduction of the $M$In$_2$ plane increases the in-plane interaction of the Ce ions and this may result in a more two-dimensional (2D) character of the systems what may be responsible for the strong enhancement of the superconducting transition temperature $T_c$ in the Ce 115 compounds in comparison to CeIn$_3$ \cite{Monthoux2001,Monthoux2002}. In this article we will review the pressure--temperature-magnetic field phase diagram of CeRhIn$_5$ and discuss the effect of Sn doping in detail. 

\section{Phase Diagram of CeRhIn$_5$} 

\begin{figure}[t]
\begin{center}
\includegraphics{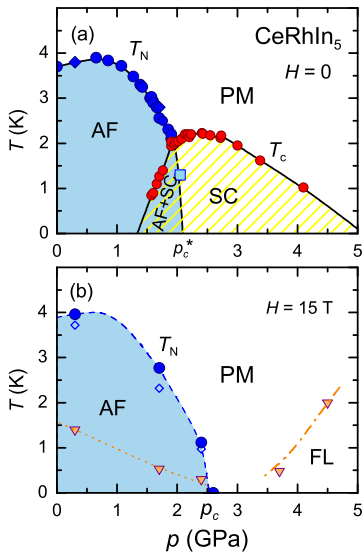}
\end{center}
\caption{(Color online) (a) High pressure phase diagram of CeRhIn$_5$ at zero magnetic field. Antiferromagnetism (AF) is suppressed at $p_c^\star$. Above $p_c^\star$ CeRhIn$_5$ is superconducting (SC). A coexistence regime AF+SC exists below $p_c^\star$. (b) Phase diagram under high magnetic field $H = 15$ T when superconductivity is suppressed. Triangles mark the upper limit of a Fermi liquid regime (FL) determined from resistivity measurements. 
}
\label{f1}
\end{figure}

CeRhIn$_5$ turns out to be an ideal system to study the interplay of magnetism and superconductivity. At ambient pressure it orders antiferromagnetically below $T_N = 3.8$~K with an incommensurate wave vector \boldmath $Q_{\rm ic}$ \unboldmath $=(0.5, 0.5, 0.298) $\cite{Bao2001,Raymond2007}. The pressure--temperature $(p,T)$ phase diagram has been studied previously in detail by different groups and nice consensus has been observed on the phase diagram (see e.g. Ref.~\citen{Knebel2010}). Under application of high pressure, $T_N (p)$ has a smooth maximum for $p=0.7$~GPa and is monotonously suppressed for higher pressure (see Fig. \ref{f1}a)). At the first critical pressure $p_c^\star \approx 2$~GPa, when $T_N = T_c$, antiferromagnetism is rapidly suppressed. Below this pressure a small domain of coexistence of magnetism and superconductivity (AF +SC) exists. Microscopic evidence for this coexistence was given by NQR experiments under high pressure.\cite{Yashima2009} It was shown that the magnetic structure gets commensurable with \boldmath $Q_{\rm af}$ \unboldmath $=(0.5, 0.5, 0.5)$ when entering in the coexisting phase AF +SC at $p \sim 1.7$~GPa. Furthermore, the spin-lattice relaxation $1/T_1$ is homogeneous and independent on the local site what is a nice hint of the coexistence of both states below $p_c^\star$ \cite{Kawasaki2003,Yashima2007,Yashima2009}. The results from neutron diffraction experiments is less clear. Here a continuous change of the incommensurability with pressure has been observed as function of pressure, but no decisive conclusion of the ordering vector in the AF+SC state could be given \cite{Raymond2008,Aso2009}. 

\begin{figure}[t]
\begin{center}
\includegraphics{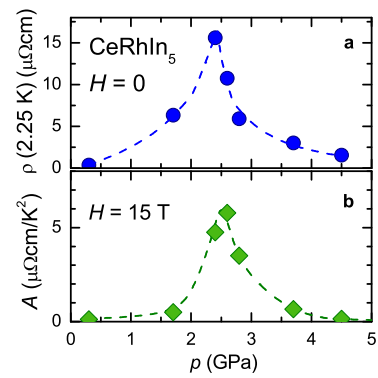}
\end{center}
\caption{(Color online) Pressure dependence of a) the resistivity at $T= 2.25$~K just above the superconducting transition, b) the $A$ coefficient of the resistivity measured at a field $H=15$~T far above the upper critical field $H_{c2}$. Both quantities are strongly enhanced around the critical pressure $p_c$. (Lines are guides for the eye.)
}
\label{f2}
\end{figure}

The superconducting transition temperature has a maximum at $p_c = 2.5$ GPa. This pressure corresponds to a linear extrapolation of $T_N (p)$ to zero and would be the critical pressure of the system in absence of superconductivity. Indications of the location of  the critical pressure are shown in Fig. \ref{f2}. The resistivity $\rho$ at $T=2.2$ K just above the onset of superconductivity  shows a strong maximum as function of pressure at $p_c$ due to the enhancement of the scattering caused by critical fluctuations \cite{Miyake2002}. 
The resistivity as function of temperature in this pressure range shows clearly a non-Fermi liquid behavior with a sub-linear  dependence above the superconducting transition \cite{Knebel2008, Park2008d}. Several theoretical proposals for a $T-$ linear dependence are discussed in the literature such as two-dimensional spin fluctuations, critical valence fluctuations or a selective Mott transition.\cite{Moriya2003, Watanabe2010, Pepin2008}. However, the deeper meaning of this very unconventional temperature dependence is not completely understood and, at least in our view, the sub-linearity is much more an effect of a cross-over to the low temperature regime in difference to Ref.~\citen{Park2008d}.

Another important indicator of the critical pressure can be obtained by studying the phase diagram of CeRhIn$_5$ under magnetic fields above the superconducting upper critical field $H_{c2}$ as shown in Fig.\ref{f1} b. For $H = 15$ T, antiferromagnetism is suppressed at $p_c = 2.5$ GPa and concomitant the temperature region where the resistivity follows a nice $T^2$ dependence collapses. Only far above the critical pressure a nice $T^2$ dependence of the resistivity is recovered. The  $A$ coefficient $A = (\rho(T)- \rho_0)/T^2$ for $T \to 0$ at $H = 15$ T is shown in Fig.\ref{f2} b) \cite{Knebel2008}. In difference to the prototypical heavy-fermion superconductor CeCu$_2$Si$_2$ where the magnetic quantum critical point ($p= 0$) and a sharp valence transition (at $p=4$~GPa) are clearly separated in pressure and two different superconducting domes exist \cite{Yuan2003, Holmes2004}, in most other heavy fermion superconductors such a clear separation of both criticalities is not possible. Thus, as function of pressure both criticalities seem to fall together, e.g.~in CeIn$_3$ \cite{Knebel2001}, CePd$_2$Si$_2$ \cite{Demuer2001}, and also here in CeRhIn$_5$ only one maximum in the pressure dependence of the $A$ coefficient occurs. The disappearance of magnetism at $p_c$ appears to be also driven by the valence fluctuations. Valence and spin fluctuations are here strongly coupled.

\begin{figure}[t]
\begin{center}
\includegraphics{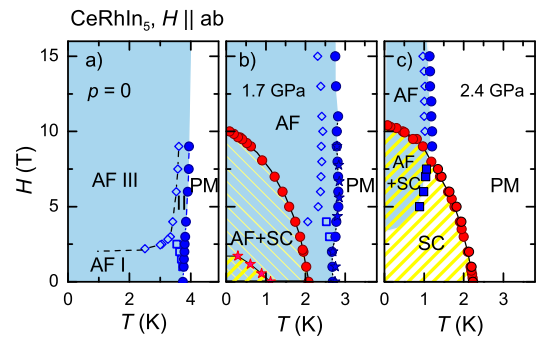}
\end{center}
\caption{(Color online) Magnetic field phase diagram of CeRhIn$_5$ at different pressures for a magnetic field applied in the $ab$ plane (blue symbols for magnetic transitions, red symbols for superconductivity). (a-b) The magnetic phase diagram is almost unchanged compared to $p=0$ up to $p_c^\star$. However, $H_{c2} (T)$ detected by specific heat (red stars, taken from Ref.\citen{Park2008b}) is much lower than that detected by resistivity. This indicates the inhomogeneous superconducting state observed below $p_c^\star$. Remarkably magnetic order is induced inside the superconducting dome in the pressures $p_c^\star < p < p_c$, as shown for $p=2.4$~GPa in (c). For $p> p_c$ no magnetic order appears.)
}
\label{f3}
\end{figure}

The application of a magnetic field will lead to another test of the interplay between antiferromagnetism and superconductivity. The magnetic phase diagram of CeRhIn$_5$ has been studied under magnetic fields up to 60 T by magnetization measurements.  In magnetization measurements for $H \perp c$ two transitions have been observed at low temperatures at $H_{ic} \approx 2$~T and $H _m\approx 50$~T \cite{Takeuchi2001}. The low temperature magnetic phase diagram has been established by specific heat measurements showing three different magnetic phases \cite{Cornelius2001} and their magnetic structures have been determined by neutron scattering \cite{Raymond2007}. Figure~\ref{f3}a) shows the $H-T$ phase diagrams at ambient pressure. The incommensurate structure (AF I) changes to commensurate (AF III) for a field of $H_{ic}\approx 2.25$~T with an ordering vector \boldmath $Q$\unboldmath $=(0.5, 0.5, 0.25)$ and the magnetic structure has changed to a collinear one. The ordering in phase AF II is also incommensurate \cite{Raymond2007}. 
	
Figure \ref{f3}b) shows the magnetic phase diagram for $p = 1.7$ GPa. From the specific heat experiments the phase diagram is similar to that at ambient pressure with three different antiferromagnetic structures. However, the structure of these phases has not been determined up to now. On top of the magnetic phase diagram is set the superconducting phase diagram. Here a strong difference between $H_{c2} (T)$ determined by resistivity and specific heat has been observed \cite{Park2008b}. Furthermore, it should be noted that, as discussed above, from the NQR spectra it has been concluded that the magnetic structure is commensurate with \boldmath $Q_{\rm af}$\unboldmath $=(0.5, 0.5, 0.5)$ in the AF+SC coexistence regime \cite{Yashima2007}. Thus, the complete determination of the magnetic structure of the different states is still open. 
	
The most exciting case appears in the pressure range $p_c^\star < p < p_c$.\cite{Knebel2006} The fascinating observation is that magnetism is induced deep inside the superconducting phase by applying a magnetic field.\cite{Knebel2006, Park2006} The microscopic nature of the field induced state has not been determined yet. An appealing pictures is that magnetism will appear initially from the vortex cores. As the system is in that peculiar pressure region very close to a magnetic instability the magnetic correlation length will be very long (compared to the atomic length scale) and thus long range correlations will appear when under magnetic field the vortex cores will approach each other.\cite{Zhang2002, Demler2004} For pressure $p>p_c$ no indication of re-entrant antiferromagnetism is observed. The collapse of antiferromagnetism induced under magnetic field coincided with the critical pressure $p_c$.

We want to emphasize that $p_c$ coincides with an abrupt change of the Fermi surface \cite{Shishido2005}.  Below $p_c$ the Fermi surface of CeRhIn$_5$ is approximately the same as those of the non $f$ compound LaRhIn$_5$, but above $p_c$ it has the same form and volume than that of CeCoIn$_5$. This change of the Fermi surface has been interpreted as a transition from an $4f$ localized to a $4f$ itinerant state. However, this interpretation is still under debate and this change of the Fermi surface may be caused by the interplay of magnetism and critical valence fluctuations \cite{Watanabe2010} or some local quantum criticality at $p_c$ \cite{Harrison2004, Shishido2005, Park2008d}.
 
No microscopic experiments exist to study this new field induced state up to now. This is mainly due to the difficulty to realize the triple extreme conditions in neutron scattering experiments, low temperature, hydrostatic pressure $p > 2$ GPa, and high magnetic field.

\section{Superconductivity}

\begin{figure}
\begin{center}
\includegraphics{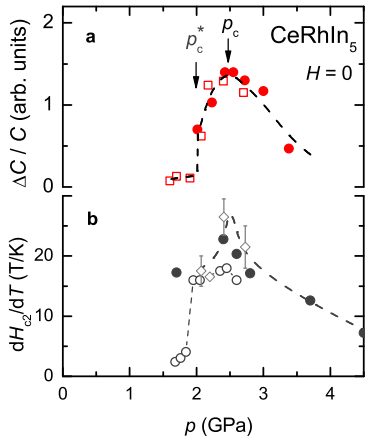}
\end{center}
\caption{(Color Alaine) (a) Specific heat jump $\Delta C/C$ at the superconducting transition as function of pressure. (Different symbols corresponds to different experiments. (b) Pressure dependence of the initial slope of the upper critical field at the superconducting transition. Full circles are from the resistivity experiments \cite{Knebel2008}, open diamonds are from ac calorimetry \cite{Knebel2006},  open circles are taken from Ref.~\citen{Park2008b}. (The pressure of Ref.~\citen{Park2008b} has been normalized to our experiments.)}
\label{deltaC}
\end{figure}

Two remarkable features appear in the pressure phase diagram of CeRhIn$_5$ : (i) the abrupt suppression of the antiferromagnetic order when superconductivity sets in above $T_N$ at the pressure $p_c^\star$ and (ii) the smooth pressure dependence of $T_c (p)$ through $p_c$.

Figure \ref{deltaC} a) shows the pressure dependence of the normalized specific heat anomaly $\Delta C/C$ at the superconducting transition in zero magnetic field. Two points get obviously: firstly, the superconducting anomaly in the AF+SC state below $p_c^\star$ the specific heat anomaly at the superconducting transition is very small and increases step-like at $p_c^\star$. Secondly, at the critical pressure $p_c$ 
the specific heat jump shows a pronounced maximum and decreases to higher pressures. For a conventional BCS-like superconductor in the weak coupling limit, the jump at the superconducting transition has an universal value, independent of $T_c$. The fact that the specific heat anomaly below $p_c^\star$ is very small has been first interpreted as indication that superconductivity may not be bulk, but only a small volume fraction shows bulk superconductivity \cite{Knebel2004}. As mentioned above now the interpretations is different as the coexistence of antiferromagnetism and superconductivity has been shown explicitly in the NQR experiments \cite{Yashima2007}.  
However, the pressure evolution of the specific heat jump at $H =0$ indicates that 
the onset of superconductivity must be accompanied with a change in the magnetic properties leading to a non BCS anomaly.
antiferromagnetism and superconductivity are competing phenomena on the Fermi surface and they are intuitively coupled. 

Further insights on the superconducting properties can be obtained from the temperature dependence of the upper critical field. From the initial slope it should be possible to estimate the effective mass of the charge carriers. The pressure dependence of the initial slope is shown in Fig. \ref{deltaC}b). In the AF+SC state the initial slope as detected in the ac calorimetry \cite{Park2008b} increases gradually with pressure up to $p_c^\star$, but at $p_c^\star$ a step increase of the initial slope appears. Thus in low magnetic fields an abrupt change in the electronic properties must occur at $p_c^\star$ at the border to the pure superconducting state. Looking at the properties of the Fermi surface this abrupt increase can be interpreted in that way that the average Fermi velocity decreases strongly or in terms of the effective mass of the charge carriers a jump of the average effective mass which is connected to the change of the Fermi surface for $H = 0$ at $p_c^\star$. At the hidden magnetic  quantum critical point at $p_c \approx 2.5$ GPa the initial slope has a rather pronounced maximum and corresponds to a maximum in the effective mass  as has been already discussed for the $A$ coefficient. This reflects itself again in the superconducting properties. To analyze completely the upper critical field curve, at least above $p_c^\star$, it is necessary to take into account strong coupling effects. For CeRhIn$_5$ we found a strong coupling parameter of $\lambda = 2.2$ close to the critical pressure, thus the mass enhancement due to the superconducting coupling is $ m_{sc}^\star/{m_b} = \lambda +1 \approx 3.2$ \cite{Knebel2008}. However, compared to the pressure variation of the effective mass as determined from the transport measurements at high pressure this is a rather smooth variation.  
De facto the strong coupling constant changes under pressure but its changing is not directly linked to the pressure variation of $m^\star$ as $m_b$ itself can change drastically under pressure notably near a valence transition.

There is no conclusive determination of the superconducting order parameter of the coexistence regime for $p<p_c^\star$. In the NQR experiments a residual density of states is observed \cite{Kawasaki2003}. Evidence for $d$ wave superconductivity has been reported from angular dependent specific heat experiments \cite{Park2008c}. Above $p_c^\star$ it is well accepted in the literature that CeRhIn$_5$ is a $d$-wave superconductor with line nodes as the non magnetic compounds CeCoIn$_5$ and CeIrIn$_5$. This is mainly concluded from the $T^3$ temperature dependence of the nuclear spin relaxation rate $1/T_1$and the absence of the Hebel-Slichter peak \cite{Mito2001}. 

%%%%%%%%%%%%%%%%%%%%%%%%%%%%%%%%%%%%%%%%%%%%%%%%%%%%%%%%%%%%%%%%%%

\section{Suppression of AF in CeRhIn$_5$ by Sn doping}

Another way to suppress the antiferromagnetism in CeRhIn$_5$ is given by doping. Different experiments have been performed in the past like diluting the Kondo lattice by the substitution of Ce by La \cite{Kim2002, Light2004} or an isoelectronic substitution of Rh by Co or Ir, which leads to a coexistence of antiferromagnetism and superconductivity in a broad concentration range,\cite{Pagliuso2002,Zapf2001,Christianson2005,Ohira2007,Goh2008}.   The substitutions on the transition metal site leads to rather complicated phase diagrams with a strong interplay between the incommensurate \boldmath $Q_{\rm ic}$ \unboldmath $=(0.5, 0.5, L)$ and the commensurate antiferromagnetic ordering vector \boldmath $Q_{\rm af}$ \unboldmath $ = (0.5, 0.5, 0.5)$ and superconductivity. Another possibility is given by Sn, Cd, or Hg doping on the In site \cite{Bauer2006b,Pham2006,Bauer2008} which is non-isoelectronic: Sn has one $p$ electron more  and Cd, Hg one less electron than In. Thus the effect is opposite for these substitutions. In the case of Cd doping in CeCoIn$_5$ detailed NQR experiments have inferred that the Cd ions nucleate magnetic order on a local scale by creating magnetic droplet \cite{Urbano2007} whereas the electron doping with Sn suppresses superconductivity linearly \cite{Bauer2005}. 

\begin{figure}
\begin{center}
\includegraphics{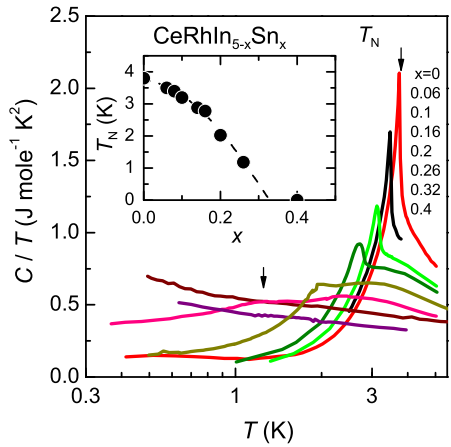}
\end{center}
\caption{(Color online) Temperature dependence of the specific heat $C$ divided by temperature $T$ of CeRhIn$_{5-x}$Sn$_x$ for different concentrations $x$ on a semi-logarithmic scale. Arrows indicate $T_N$ for $x=0$ and $x=0.26$. The inset shows the variation of the antiferromagnetic transition temperature $T_N$ $vs.$ Sn concentration $x$ (dashed line to guide the eye).}
\label{f4}
\end{figure}

In CeRhIn$_5$, Sn doping suppresses the antiferromagnetic order \cite{Bauer2006b} as the additional $p$ electron of  Sn atoms increases the conduction electron density of states and  thus the Kondo temperature will be enhanced. To study the magnetic phase diagram as function of Sn doping in more detail, we grew new single crystals of CeRhIn$_{5-x}$Sn$_x$ out of the In flux starting with a ratio Ce : Rh : In : Sn = 1 : 1 : 20 : $y$. In Ref. \citen{Bauer2006b} a linear relationship between the actual Sn concentration $x = 0.4  y$ in the crystal and the starting Sn ratio $y$ in the flux has been found. We took the same relation in the determination of the actual concentration throughout this article. In addition, to verify the actual concentration neutron diffraction experiments at ILL Grenoble on the thermal four-circle diffractometer D15 on a single crystal with an nominal concentration of $y=0.4$ have been performed recently. The result of the best refinement gives a slightly higher Sn concentration taking into account that the Sn does not occupy only the In(1) but also the In(2) site in the structure \cite{Ressouche2010} which is in difference to CeCoIn$_{5-x}$Sn$_x$ where the preferential occupation of Sn is the In(1) site within the "CeIn$_3$" planes \cite{Daniel2005}.
However, a more detailed structural  analysis has to be done to define the actual Sn concentration for all concentrations. 

Figure \ref{f4} shows the specific heat $C/T$  of CeRhIn$_{5-x}$Sn$_x$ for various Sn concentrations. With increasing $x$ the lambda-like magnetic anomaly at $T_N$ is suppressed and the magnetic transition gets smeared out. For $x = 0.26$ e.g. only a small hump is visible at the magnetic transition. A broad Schottky-like peak in $C/T$ appears around $T \approx 4$ K for $x = 0.1$ and this anomaly shifts to lower temperature with increasing $x$. In pure CeRhIn$_5$ this anomaly cannot be identified as its maximum coincides almost with $T_N$.  In CeRhIn$_5$ the entropy associated with the magnetic transition is only $0.3R \ln 2$ and the remaining entropy  $0.7R \ln 2$ is recovered only at $T=20$ K  indicating strong magnetic correlations above $T_N$. The Schottky like anomaly has  been attributed to short range magnetic correlations \cite{Bauer2006b,Bao2002}. For $x = 0.32$ the specific heat $C/T$ increases logarithmically to low temperatures in the range from at least 5 K to 0.75 K and increases stronger to lower temperatures.  This increase may be due to the onset of magnetic order at even lower temperatures. For $x = 0.4$ we find roughly $C/T \propto -\log T$ in all measured temperature range. This $-\log T$ dependence is generally found close to an antiferromagnetic quantum critical instability in a 3D spin-density scenario at intermediate temperatures followed by a square root temperature dependence at lowest temperatures.  However, more detailed very low temperature specific heat experiments are required to study this criticality. 

%It should be noted that even for much higher nominal Sn concentrations $C/T$ seems shows a strong increase to lower temperatures (not shown), but the actual Sn concentration $x$ for these crystals has not been determined up to now. 
The inset in Fig. \ref{f4} shows the the N\'eel temperature as function of the actual Sn concentration $x$  and we find good agreement of $T_N$ $vs.$ $x$  with previous published data \cite{Bauer2006b}. A linear extrapolation of $T_N (x)$ indicates a critical concentration of  $x_c \approx 0.35$.  However, low temperature thermal-expansion measurements suggest higher critical concentration of $x_c \approx 0.46$ as $T_N (x)$ deviates at low temperatures from the linear extrapolation and this may be due to the weak disorder induced by doping.\cite{Donath2009} A similar tail of $T_N(x)$ has been observed e.g. in CeIn$_{3-x}$Sn$_x$ series.\cite{Pedrazzini2004}

\begin{figure}
\begin{center}
\includegraphics{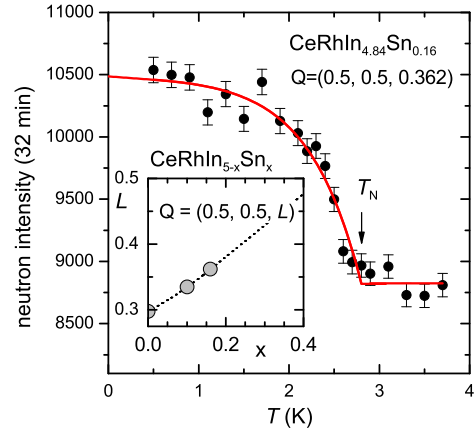}
\end{center}
\caption{(Color online) Temperature dependence of the peak intensity measured for $Q = (0.5, 0.5, 0.362)$ for CeRhIn$_{4.84}$Sn$_{0.16}$).  The inset shows the development of the magnetic ordering vector as function of Sn concentration. Lines are to guide the eye.  

}
\label{f5}
\end{figure}

The magnetic structure of Sn doped single crystals have been studied at ILL Grenoble on diffractometer D23. Details of the experiments will be published in a forthcoming article \cite{Ressouche2010}. Sn doping leads to a continuous change of the magnetic propagation vector. 
In Fig. \ref{f5} we show the temperature dependence of the peak intensity measured at the ordering wave vector \boldmath $Q$ \unboldmath $= (0.5, 0.5, 0,362)$ for CeRhIn$_{4.84}$Sn$_{0.16}$. The magnetic moment $\mu = 0.49(1) \mu_B$ is reduced compared to the moment $\mu = 0.59(1) \mu_B$ of the pure compound.\cite{Raymond2007}
 The N\'eel temperature $T_N = 2.8$ K determined from the neutron diffraction is in good agreement with the specific heat experiment. The inset in Fig. \ref{f3} shows the evolution of the  component of the propagation vector along the $c$ direction as function of Sn doping $x$. With increasing Sn contents it changes continuously from $L = 0.298$ for $x=0$ to $L = 0.362$ at $x=0.16$. Interestingly, a rough extrapolation to the critical concentration $x_c\approx 0.35$ would indicate a commensurate magnetic structure very close to the critical concentration. 

\begin{figure}
\begin{center}
\includegraphics{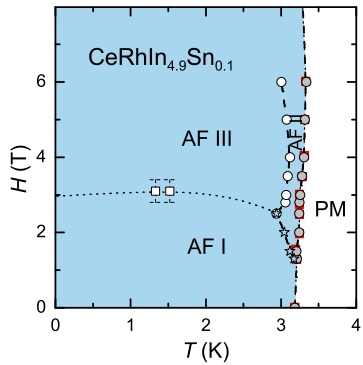}
\end{center}
\caption{(Color online) Magnetic phase diagram of CeRhIn$_{4.9}$Sn$_{0.1}$  for a magnetic field $H \parallel a$ in the basal $ab$ plane from specific heat experiments. In analogy to pure CeRhIn$_5$ it shows three different antiferromagnetic phases under magnetic field.}
\label{f6}
\end{figure}

Next we want to focus on the phase diagram under magnetic field $H \parallel a$ axis which is plotted in Fig. \ref{f6} for a Sn concentration of $x = 0.1$. The phase diagram has been obtained by specific heat experiments under magnetic field. Again, as in the pure compound, three different magnetic phases can be assigned. We know from neutron scattering that at zero field the magnetic structure is incommensurable with a component along the $c$ direction of $L=0.335$. In analogy to CeRhIn$_5$, under magnetic field the structure may change at a field of $H = 3$ T probably to an commensurate structure with an ordering vector \boldmath $Q$ \unboldmath $= (1/2, 1/2, 1/4)$. 

The important point we want to stress is that the effect on the magnetic properties induced by Sn doping or hydrostatic pressure is very similar. However, a significant difference is that superconductivity is not observed at ambient pressure.  In a more general view, antiferromagnetism and superconductivity are strongly interacting order parameters on the Fermi surface. In the pure CeRhIn$_5$ the pressure induced change of the Fermi leads close to 2 GPa implies a suppression of antiferromagnetism and the formation of a pure superconducting ground state. Sn doping however does not only modify the Fermi surface properties, but also induces an additional scattering mechanism which acts as pair breaking. Detailed studies of this effect have been performed for the suppression of superconductivity in CeCoIn$_5$ by Sn doping \cite{Bauer2005, Daniel2005}.

It should be noted that pressure induced superconductivity has been observed in CeRhIn$_{4.84}$Sn$_{0.16}$ in resistivity and specific heat experiments.\cite{Mendonca2008} The obtained pressure--temperature phase diagram for that Sn concentration $x=0.16$ is very similar to the pure compound, but shifted to lower pressures.

\section{Conclusions and Perspectives}

\begin{figure}
\begin{center}
\includegraphics[width=.9\hsize,clip]{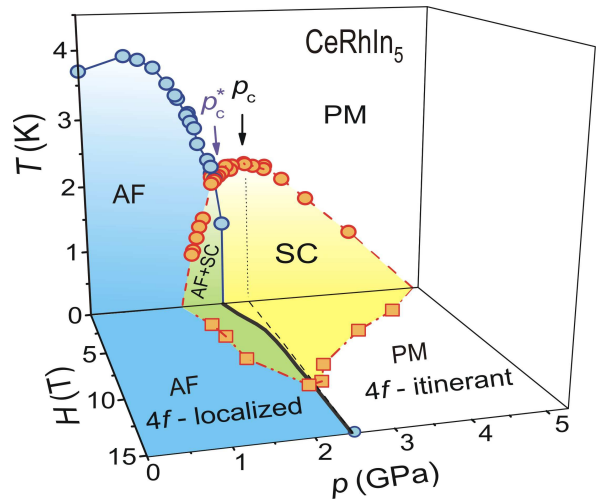}
\end{center}
\caption{(Color online) Combined temperature, pressure and field $H\perp c$ phase diagram of CeRhIn$_5$  with antiferromagnetic (blue), superconducting (yellow), and coexistence AF+SC (green) phases. The thick black line in the $H-p$ plane indicates the proposed line where the Fermi surface changes from 4$f$ "localized" (small Fermi surface and topology comparable to LaRhIn$_5$), to 4$f$ "itinerant" (large Fermi surface as in CeCoIn$_5$).}
\label{3D}
\end{figure}

CeRhIn$_5$ is an exciting heavy-fermion system which allows to study in detail the interplay of long range antiferromagnetic order and superconductivity. Due to the fact that in this system the transition temperatures of antiferromagnetism and superconductivity are of the same order of magnitude a precise determination of the high pressure phase diagram could be realized. The combined $p, H, T$ phase diagram is plotted in Fig. \ref{3D}. Under pressure and magnetic field three different regions appear, purely antiferromagnetic AF, a coexistence regime AF+SC, and a purely superconducting phase SC above $p_c^\star$. Thus as function of pressure at zero field the series of phases is AF -- AF+SC -- SC. 
In the pressure region between $p_c^\star$ and $p_c$ under application of magnetic field magnetic order is induced and one can observe the cascades of phases SC -- AF+SC -- AF with increasing field. There seems to be a difference between the pressure induced  AF+SC  and field induced AF+SC  states. When the AF state is established first under pressure at $H =0$, the measured specific heat jump and the initial slope of the upper critical field are still small and the superconducting transition at $T_c$ may coincide with a magnetic transition from an incommensurate to a commensurate structure. In contrast, when above $p_c^\star$ first superconductivity appears on cooling, the field induced AF state seems to have no influence on the SC properties such as $H_{c2} (T)$. Up to now no microscopic measurements exist in the pressure range from $p_c^\star$ to $p_c$ under magnetic field, obviously the field induced magnetic state is directly connected to the appearance of vortices in the mixed state of the superconductor. 

The substitution of In by Sn suppresses the magnetic order of CeRhIn$_5$. While the magnetic structure as function of Sn doping and hydrostatic pressure seems to have a very similar behavior, superconductivity at ambient pressure appears only under pressure what indicates that Sn doping suppresses superconductivity.

\section*{Acknowledgment}

This work has been financially supported by the French ANR programs DELICE and CORMAT.

%\begin{thebibliography}{9}
%\bibitem{jpsj} The abbreviation for JPSJ must be ``J. Phys. Soc. Jpn." in the reference list.
%\bibitem{instructions} More abbreviations of journal titles are listed in ``Instructions for Preparation of Manuscript".
%\end{thebibliography}

%\bibliographystyle{jpsj}
%\bibliography{D:/bibtex/Ce_references}

\begin{thebibliography}{10}
\providecommand{\url}[1]{\texttt{#1}}
\providecommand{\urlprefix}{URL }
\providecommand{\eprint}[2][]{\url{#2}}

\bibitem{Mathur1998}
N.~D. Mathur, F.~M. Grosche, S.~R. Julian, I.~R. Walker, D.~M. Freye, R.~K.~W.
  Haselwimmer and G.~G. Lonzarich: Nature \textbf{394} (1998) 39.

\bibitem{Sachdev2010}
S.~Sachdev: phys. stat. sol. B \textbf{247} (2010) 537.

\bibitem{Bourbonnais2008}
C.~Bourbonnais and D.~J\'erome: \emph{Physics of Organic Superconductors and
  Conductors, {\rm SpringerSeries in Material Science Vol.110}} (Springer,
  Berlin, 2008) p. 357.

\bibitem{Paglione2010}
J.~Paglione and R.~L. Greene: Nature Phys. \textbf{6} (2010) 645.

\bibitem{Knebel2001}
G.~Knebel, D.~Braithwaite, P.~C. Canfield, G.~Lapertot and J.~Flouquet: Phys.
  Rev. B \textbf{65} (2001) 024425.

\bibitem{Monthoux2001}
P.~Monthoux and G.~G. Lonzarich: Phys. Rev. B \textbf{63} (2001) 054529.

\bibitem{Monthoux2002}
P.~Monthoux and G.~G. Lonzarich: Phys. Rev. B \textbf{66} (2002) 224504.

\bibitem{Bao2001}
W.~Bao, P.~G. Pagliuso, J.~L. Sarrao, J.~D. Thompson, Z.~Fisk and J.~W. Lynn:
  Phys. Rev. B \textbf{64} (2001) 020401.

\bibitem{Raymond2007}
S.~Raymond, E.~Ressouche, G.~Knebel, D.~Aoki and J.~Flouquet: J. Phys.:
  Condens. Matter \textbf{19} (2007) 242204.

\bibitem{Knebel2010}
G.~Knebel, D.~Aoki, J.-P. Brison, L.~Howald, G.~Lapertot, J.~Panarin,
  S.~Raymond and J.~Flouquet: phys. stat. sol. B \textbf{{247}} (2010) 557.
  

\bibitem{Yashima2009}
M.~Yashima, H.~Mukuda, Y.~Kitaoka, H.~Shishido, R.~Settai and Y.~Onuki: Phys.
  Rev. B \textbf{79} (2009) 214528.

\bibitem{Kawasaki2003}
S.~Kawasaki, T.~Mito, Y.~Kawasaki, G.~q~Zheng, Y.~Kitaoka, D.~Aoki, Y.~Haga and
  \=Onuki: Phys. Rev. Lett. \textbf{91} (2003) 137001.

\bibitem{Yashima2007}
M.~Yashima, S.~Kawasaki, H.~Mukuda, Y.~Kitaoka, H.~Shishido, R.~Settai and
  Y.~\=Onuki: Phys. Rev. B \textbf{76} (2007) 020509.

\bibitem{Raymond2008}
S.~Raymond, G.~Knebel, D.~Aoki and J.~Flouquet: Phys. Rev. B \textbf{77} (2008)
  172502.

\bibitem{Aso2009}
N.~Aso, K.~Ishii, H.~Yoshizawa, T.~Fujiwara, Y.~Uwatoko, G.-F. Chen, N.~K. Sato
  and K.~Miyake: Journal of the Physical Society of Japan \textbf{78} (2009)
  073703.

\bibitem{Miyake2002}
K.~Miyake and O.~Narikiyo: J. Phys. Soc. Jpn. \textbf{71} (2002) 867.

\bibitem{Knebel2008}
G.~Knebel, D.~Aoki, J.-P. Brison and J.~Flouquet: Journal of the Physical
  Society of Japan \textbf{77} (2008) 114704.

\bibitem{Park2008d}
V.~A. Park, T. amd~Sidorov, F.~Ronning, J.-X. Zhu, Y.~Tokiwa, H.~Lee, E.~D.
  Bauer, R.~Movshovich, J.~L. Sarrao and J.~D. Thompson: Nature \textbf{456}
  (2008) 366.

\bibitem{Moriya2003}
T.~Moriya: Rep. Prog. Phys. \textbf{66} (2003) 1299.

\bibitem{Watanabe2010}
S.~Watanabe and K.~Miyake: arXiv: 1009.5757v1 .

\bibitem{Pepin2008}
C.~P\'{e}pin: Physical Review B (Condensed Matter and Materials Physics)
  \textbf{77} 245129.

\bibitem{Yuan2003}
H.~Q. Yuan, F.~M. Grosche, M.~Deppe, C.~Geibel, G.~Sparn and F.~Steglich:
  Science \textbf{302} (2003) 2104.

\bibitem{Holmes2004}
A.~T. Holmes, D.~Jaccard and K.~Miyake: Phys. Rev. B \textbf{69} (2004) 024508.

\bibitem{Demuer2001}
A.~Demuer, D.~Jaccard, I.~Sheikin, S.~Raymond, B.~Salce, J.~Thomasson,
  D.~Braithwaite and J.~Flouquet: J. Phys.: Condens. Matter \textbf{13} (2001)
  9335.

\bibitem{Park2008b}
T.~Park, M.~J. Graf, L.~Boulaevskii, J.~L. Sarrao and J.~D. Thompson: Proc.
  Natl. Acad. Sci. U.S.A. \textbf{105} (2008) 6825.

\bibitem{Takeuchi2001}
T.~Takeuchi, T.~Inoue, K.~Sugiyama, D.~Aoki, Y.~Tokiwa, Y.~H.~K. Kindo and
  Y.~\=Onuki: J. Phys. Soc. Jpn. \textbf{70} (2001) 877.

\bibitem{Cornelius2001}
A.~L. Cornelius, P.~G. Pagliuso, M.~F. Hundley and J.~L. Sarrao: Phys. Rev. B
  \textbf{64} (2001) 144411.

\bibitem{Knebel2006}
G.~Knebel, D.~Aoki, D.~Braithwaite, B.~Salce and J.~Flouquet: Phys. Rev. B
  \textbf{74} (2006) 020501(R).

\bibitem{Park2006}
T.~Park, F.~Ronning, H.~Q. Yuan, M.~B. Salamon, R.~Movshovich, J.~L. Sarrao and
  J.~D. Thompson: Nature \textbf{440} (2006) 65.

\bibitem{Zhang2002}
Y.~Zhang, E.~Demler and S.~Sachdev: Phys. Rev. B \textbf{66} (2002) 094501.

\bibitem{Demler2004}
E.~Demler, W.~Hanke and S.-C. Zhang: Rev. Mod. Phys. \textbf{76} (2004) 909.

\bibitem{Shishido2005}
H.~Shishido, R.~Settai, H.~Harima and Y.~\=Onuki: J. Phys. Soc. Jpn.
  \textbf{74} (2005) 1103.

\bibitem{Harrison2004}
N.~Harrison, U.~Alver, R.~G. Goodrich, I.~Vekhter, J.~L. Sarrao, P.~G.
  Pagliuso, N.~O. Moreno, L.~Balicas, Z.~Fisk, D.~Hall, R.~T. Macaluso and
  J.~Y. Chan: Phys. Rev. Lett. \textbf{93} (2004) 186405.

\bibitem{Knebel2004}
G.~Knebel, M.~A. Measson, B.~Salce, D.~Aoki, D.~Braithwaite, J.~P. Brison and
  J.~Flouquet: J. Phys.: Condens. Matter \textbf{16} (2004) 8905.

\bibitem{Park2008c}
T.~Park, E.~D. Bauer and J.~D. Thompson: Phys. Rev. Lett. \textbf{101} 177002
  (pages~4).

\bibitem{Mito2001}
T.~Mito, S.~Kawasaki, G.~q~Zheng, Y.~Kawasaki, K.~Ishida, Y.~Kitaoka, D.~Aoki,
  Y.~Haga and \=Onuki: Phys. Rev. B \textbf{63} (2001) 220507(R).

\bibitem{Kim2002}
J.~S. Kim, J.~Alwood, D.~Mixson, P.~Watts and G.~R. Stewart: Phys. Rev. B
  \textbf{66} (2002) 134418.

\bibitem{Light2004}
B.~E. Light, R.~S. Kumar, A.~L. Cornelius, P.~G. Pagliuso and J.~L. Sarrao:
  Phys. Rev. B \textbf{69} (2004) 024419.

\bibitem{Pagliuso2002}
P.~G. Pagliuso, R.~Movshovich, A.~D. Bianchi, M.~Nicklas, N.~O. Moreno, J.~D.
  Thompson, M.~F. Hundley, J.~L. Sarrao and Z.~Fisk: Physica B \textbf{312}
  (2002) 129.

\bibitem{Zapf2001}
V.~S. Zapf, E.~J. Freeman, E.~D. Bauer, J.~Petricka, C.~Sirvent, N.~A.
  Frederick, R.~P. Dickey and M.~B. Maple: Phys. Rev. B \textbf{65} (2001)
  014506.

\bibitem{Christianson2005}
A.~D. Christianson, A.~Llobet, W.~Bao, J.~S. Gardner, I.~P. Swainson, J.~W.
  Lynn, J.~M. Mignot, K.~Prokes, P.~G. Pagliuso, N.~O. Moreno, J.~L. Sarrao,
  J.~D. Thompson and A.~H. Lacerda: Phys. Rev. Lett. \textbf{95} (2005) 217002.

\bibitem{Ohira2007}
S.~Ohira-Kawamura, H.~Shishido, A.~Yoshida, R.~Okazaki, H.~Kawano-Furukawa,
  T.~Shibauchi, H.~Harima and Y.~Matsuda: Phys. Rev. B \textbf{76} (2007)
  132507.

\bibitem{Goh2008}
S.~K. Goh, J.~Paglione, M.~Sutherland, E.~C.~T. O'Farrell, C.~Bergemann, T.~A.
  Sayles and M.~B. Maple: Physical Review Letters \textbf{101} 056402.

\bibitem{Bauer2006b}
E.~Bauer, D.~Mixson, F.~Ronning, N.~Hur, R.~Movshovich, J.~Thompson, J.~Sarrao,
  M.~Hundley, P.~Tobash and S.~Bobev: Physics B \textbf{378-80} (2006) 142.

\bibitem{Pham2006}
L.~D. Pham, T.~Park, S.~Maquilon, J.~D. Thompson and Z.~Fisk: Phys. Rev. Lett.
  \textbf{97} (2006) 056404.

\bibitem{Bauer2008}
E.~D. Bauer, F.~Ronning, S.~Maquilon, L.~D. Pham, J.~D. Thompson and Z.~Fisk:
  Physica B \textbf{403} (2008) 1135.

\bibitem{Urbano2007}
R.~R. Urbano, B.-L. Young, N.~J. Curro, J.~D. Thompson, L.~D. Pham and Z.~Fisk:
  Phys. Rev. Lett. \textbf{99} (2007) 146402.

\bibitem{Bauer2005}
E.~D. Bauer, C.~Capan, F.~Ronning, R.~Movshovich, J.~D. Thompson and J.~L.
  Sarrao: Phys. Rev. Lett. \textbf{94} (2005) 047001.

\bibitem{Ressouche2010}
E.~Ressouche: unpublished .

\bibitem{Daniel2005}
M.~Daniel, E.~D. Bauer, S.-W. Han, C.~H. Booth, A.~L. Cornelius, P.~G. Pagliuso
  and J.~L. Sarrao: Phys. Rev. Lett. \textbf{95} (2005) 016406.

\bibitem{Bao2002}
W.~Bao, G.~Aeppli, J.~W. Lynn, P.~G. Pagliuso, J.~L. Sarrao, M.~F. Hundley,
  J.~D. Thompson and Z.~Fisk: Phys. Rev. B \textbf{65} (2002) 100505.

\bibitem{Donath2009}
J.~G. Donath, F.~Steglich, E.~D. Bauer, F.~Ronning, J.~L. Sarrao and
  P.~Gegenwart: EPL (Europhysics Letters) \textbf{87} (2009).

\bibitem{Pedrazzini2004}
P.~Pedrazzini, M.~Berisso, N.~Caroca-Canales, M.~Deppe, C.~Geibel and
  J.~Sereni: Eur. Phys. J. B \textbf{38} (2004) 445.

\bibitem{Mendonca2008}
L.~Mendon\c{c}a Ferreira, T.~Park, V.~Sidorov,
  M.~Nicklas, E.~M. Bittar, R.~Lora-Serrano, E.~N. Hering, S.~M. Ramos, M.~B.
  Fontes, E.~Baggio-Saitovich, H.~Lee, J.~L. Sarrao, J.~D. Thompson and P.~G.
  Pagliuso: Phys. Rev. Lett. \textbf{101} (2008) 017005.

\end{thebibliography}

\end{document}